\documentclass[twocolumn,superscriptaddress,floatfix]{revtex4}

\usepackage{graphicx}

\begin{document}
\title{Dynamical instability and dispersion management\\
of an attractive condensate in an optical lattice}
\author{G. Barontini}
\affiliation{LENS and Dipartimento di Fisica, Universit\`a di Firenze via Nello
Carrara 1, I-50019 Sesto Fiorentino, Italy}
\author{M. Modugno}
\affiliation{LENS, Dipartimento di Matematica Applicata, Universit\`a di
Firenze and INFN, Italy}
\affiliation{BEC-INFM Center, Universit\`a di Trento, I-38050 Povo, Italy}

\begin{abstract}
We investigate the stability of an attractive Bose-Einstein condensate in a moving 1D optical lattice in the presence of transverse confinement. By means of a Bogoliubov linear stability analysis we find that the system is dynamically unstable for low quasimomenta and becomes stable near the band edge, in a specular fashion with respect to the repulsive case. For low interactions the instability occurs via long wavelength excitations that are not sufficient for spoiling the condensate coherence, producing instead an oscillating density pattern both in real and momentum space. This behaviour is illustrated by simulations for the expansion of the condensate in a moving lattice.
\end{abstract}


\maketitle

The occurrence of dynamical instability and the possibility for dispersion management of Bose-Einstein condensates (BECs) propagating in optical lattices have been the object of an intensive experimental and theoretical activity in recent years \cite{morsch}. 
Dynamical instability, that originates from the interplay between nonlinearity of the interatomic interaction and periodicity of the lattice, has been proved to be a major mechanism for the disruption of the superfluid flow of repulsive BECs \cite{wuniu,smerzi,pethick,menotti,scott,modugno,cataliottinjp,fallani04,campbell}.  
Such a phenomena occurs via the exponential growth of arbitrarily small fluctuation when the relative velocity between the BEC and the lattice exceeds a critical threshold, above the critical velocity for Landau instability \cite{wuniu}.
On the other side, in a regime of low interactions where nonlinear effects are negligible, it has been demonstrated that it is possible to control the dispersion of matter wave packets by tuning the velocity of the periodic lattice, in a similar fashion to the dispersion management in optics \cite{massignan,eiermann,fallani03}. This has striking effects for example on the expansion dynamics of the condensate, that can be reversed by switching from positive to negative effective mass regimes.

The realization of BECs with tunable attractive interactions \cite{lithium,cesium,39K} opens now interesting perspectives
for the investigations of the effects of dynamical instability and dispersion management in unexplored regimes. 

In this paper we show that an attractive BEC moving in a 1D optical lattice, contrarily to the repulsive case, is dynamically unstable in the first half of the Brillouin zone and can be stabilized near the band edge. Moreover, in case of weak nonlinearity and low quasimomenta the instability occurs via long wavelength excitations that do not destroy the condensate coherence, giving rise to density modulations that extend over several sites of the lattice. The effects of this interesting behaviour are discussed by comparing the predictions of the Bogoliubov linear stability analysis with the solution of a time dependent Gross-Pitaevskii (GP) model \cite{stringari} for the waveguide expansion of the condensate in the presence of the moving lattice. We show that in the unstable regime breathing-like oscillations are produced for low velocities, and that the
absence of dynamical instability for higher velocities allows for a close inspection of negative mass effects near the band edge \cite{fallani03}.

Let us start by considering an infinite cylindrical condensate of atoms of mass $m$, radially confined by the harmonic potential $V_{\rm ho} (r) = (1/2) m \omega_r^2 r^2$, 
and subjected to the one-dimensional optical lattice $V(z)=sE_R\cos^2(k z)$.
The parameter $s$ represents the height of the lattice in units of the recoil energy $E_R=\hbar^2 k^2/2m$, $k=2\pi/\lambda$ being the lattice wavenumber. 
The presence of the radial confinement makes convenient to use dimensionless radial units.
In the following, unless otherwise stated, we will express distances in units of the
harmonic oscillator length $a_r=\sqrt{\hbar/m\omega_r}$, time in units of $\omega_r^{-1}$ 
and energies in units of $\hbar\omega_r$ (keeping the same symbols in order to simplify the notation). 

The behaviour of the system is governed by the 3D GP equation for the condensate wavefunction \cite{stringari}. 
However, owing to the radial confinement it can be conveniently described also by means of an effective 1D model, the Non-Polynomial Schr\"odinger Equation (NPSE) \cite{salasnich}. 
Though onedimensional, this model includes an effective radial-to-axial coupling,
and has been shown to provide a realistic description in 
several situations \cite{salasnich,modugno,fallani04}. 
It is obtained from the GP equation by means of a factorization of condensate wavefunction in the product of a Gaussian radial component of $z$-
and $t$-dependent width, $\sigma(z,t)$, and of an axial wave
function  $\varphi(z,t)$
solution of 
\begin{equation}
i\partial_t \varphi=\left[-\frac{1}{2}\nabla_z^2+V(z)
+\frac12\left(3\sigma^2-\frac{1}{\sigma^2}\right)\right]\varphi
\label{eq:npse}
\end{equation}
with $\sigma^2(z,t)=\sqrt{1+g|\varphi(z,t)|^2}$. The coupling constant $g$ is defined as 
$g=2Na/a_r$, $a$ being the scattering length and $N$ the number of atoms in each site of the lattice. The wavefunction $\varphi$ is normalized to unity over a single period.
In the limit of vanishing lattice, $s=0$, $\varphi$
is spatially uniform and the normalization yields $|\varphi|^2=2/\lambda$.

Equation (\ref{eq:npse}) admits stationary solutions in the form of Bloch waves
$\varphi(z,t)=\phi_p(z){\rm e}^{i(\mu t-pz)}$
where $\phi_p$ has the same period of the lattice, the Bloch wave vector $p$ represents the quasimomentum of the 
condensate and is restricted to the first Brillouin zone, and 
$\mu$ is the chemical potential.
The stationary NPSE for 
$\phi_p$ becomes
\begin{eqnarray}
\left[H_p +\frac12\left(
3\sigma^2-\frac{1}{\sigma^2}\right)
\right]\phi_p=\mu \phi_p
\end{eqnarray}
with $H_p=(i\partial_z-p)^2/2 + s E_R\cos^2(\pi z/\lambda)$.
The behavior of the system is therefore characterized by three dimensionless parameters:
$\lambda$, $s$ and $g$.

\begin{figure}
\centerline{\includegraphics[width=0.6\columnwidth,clip=]{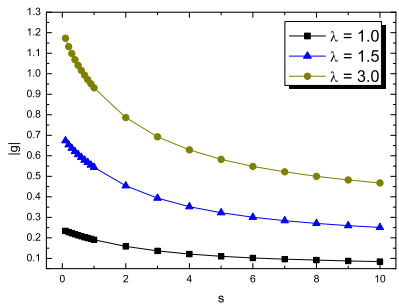}}
\caption{(Color online) Critical values $|g_{cr}|$ of the interaction strength 
as a function of $s$, for three values of $\lambda$. 
For each value of $\lambda$ the system is stable against collapse in the region 
below the corresponding line for $|g_{cr}|$.}
\label{fig:g-crit}
\end{figure}

In case of attractive interaction among the atoms, $g=-|g|$, the interaction strength
is limited by a critical value $g_{cr}$ at which the system collapses.
The value of $g_{cr}$ depends on both $\lambda$ and $s$, and its behaviour is shown in Fig. \ref{fig:g-crit}.
For $s=0$ the stability is determined by the presence of the radial trapping,
and the critical value is $|g_{cr}|=\lambda/2$. For higher values of $|g_{cr}|$
the energy functional becomes unbounded from below and the equation for $\sigma^2$ no longer admits a real solution. By increasing $s$ also the effect of the lattice becomes important and reduces the critical value of $|g|$.

Though stable against collapse the above stationary solution
can present other kind of instabilities, among which energetic (or Landau) and dynamical instability (DI). The latter can be investigated
by means of a linear analysis, 
considering time dependent perturbations of the form
$\varphi(z,t)= {\rm e}^{i(\mu t-pz)}\left[\phi_{p}(z) +\delta\phi_{p}(z,t)\right]$
where the fluctuations $\delta\phi_{p}$ can be expanded in terms 
of Bogoliubov quasiparticle amplitudes $u_{pq,n}(z)$ and $v_{pq,n}(z)$ 
with the same periodicity of the lattice
\begin{equation}
\delta\phi_{p} =\sum_{q,n}\left[u_{pq,n}{\rm e}^{i(qz-\omega_{pq,n}t)} 
+ v^*_{pq,n}{\rm e}^{-i(qz-\omega_{pq,n}t)}\right]
\end{equation}
$q$ being the quasimomentum of the excitations and $n$ the Bloch band index. 
The $u_{pq,n}$ and $v_{pq,n}$ are solutions of the Bogoliubov-like equations \cite{wuniu}
that corresponds the eigenvalue problem for the $2\times2$
matrix $M$ defined as $M\equiv(A_+ , B; -B^*, -A_-)$, with \cite{modugno}
\begin{eqnarray}
A_\pm &=& H_{p\pm q}-\mu 
+\frac12\left(\frac{9}{2}\sigma^2 -\frac{2}{\sigma^2}-
\frac{1}{2\sigma^6}\right)\\ 
B &=&
\frac{1}{4\sigma^2}\left(3 +\frac{1}{\sigma^4}\right)
g\phi_p^2\;.
\end{eqnarray}
DI occurs when the frequencies $\omega_{pq}$ of some modes
get a nonzero imaginary part, indicating an exponential growth. If initially occupied, these modes rapidly become macroscopically populated
driving the condensate away from the unperturbed state \cite{wuniu,smerzi,fallani04,modugno}. 
In a similar way, but considering the contribution $\delta E$ to the energy functional
of time independent perturbations, one can define energetic instability, occurring when $\sigma_3 M$ has negative eigenvalues ($\sigma_3$ being the third Pauli matrix) \cite{wuniu,modugno}.

The Bogoliubov frequencies $\omega_{pq}$ can be obtained by numerical diagonalization of the matrix $M$ and the corresponding dynamically unstable regions can be drawn in the $p$-$q$ plane where the imaginary component of $\omega_{pq}$ is nonvanishing.
Since complex frequencies always come in conjugate pairs, the appearance of an imaginary component corresponds to a resonance between the phonon and antiphonon branches of the spectrum \cite{wuniu}.

\begin{figure}
\centerline{\includegraphics[width=0.85\columnwidth,clip=]{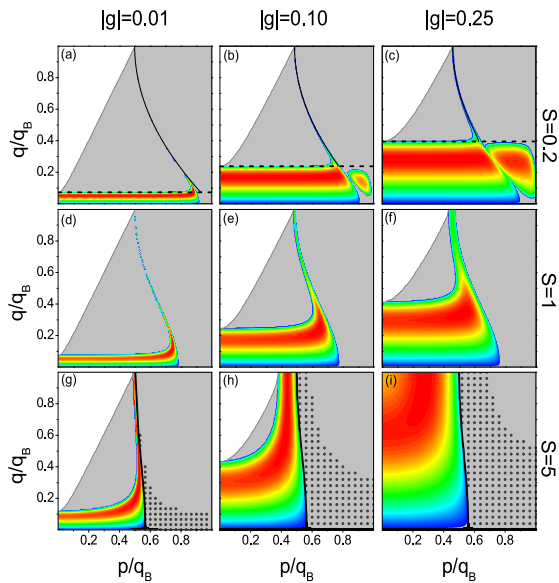}}
\caption{(Color online) Stability diagrams as a function of the quasimomenta $p$ and $q$ of the condensate and the excitations, for different values of $s$ and $g$.
Colored regions indicate where 
the system is dynamically unstable. The colour scale corresponds to the
growth rate of the unstable modes for a given $p,q$ (imaginary part of $\omega_{pq}$).
The solid and dashed lines in (a), (b) and (c) correspond
to the analytic predictions for 
$s=0$, whereas the bold line in (g), (h) and (i) refer to the case $g=0$ (see text).
Light shaded area: energetic instability. 
Dotted region in (g), (h), and (i): regimes of DI for the repulsive case $g=|g|$.
}
\label{fig:stability}
\end{figure}

Let us now focus to the case of an attractive condensate.
To illustrate a specific case we take $\lambda = 2.5$, 
but the effects we are going to discuss can be obtained in a wide range of parameters, with different values of the radial confinement and lattice periodicity.
Typical results for the stability diagrams are shown in Fig.~\ref{fig:stability} for different values of $s$ and $g$. 
In this picture we also show the regimes of energetic instability obtained from the diagonalization of $\sigma_3 M$ (shaded regions), indicating that in the attractive case the system is energetically unstable for any velocity, and therefore it is not a superfluid according to the Landau criterion \cite{stringari}.

As regards DI, these diagrams show a remarkable behaviour
compared to the case of repulsive interactions investigated so far \cite{wuniu,fallani04} (see \textit{e.g.} Fig. \ref{fig:stability}g-i), and there are at least three points that make the attractive case particularly interesting. 
i) Increasing the lattice velocity stabilizes the system:
attractive condensates are dynamically unstable at low quasimomenta, but can be stabilized above a certain critical velocity. This is just the opposite behaviour of that for repulsive BECs, that are stable below a critical velocity \cite{wuniu,fallani04}.
ii)
For low quasimomenta and shallow lattices DI takes place via long wavelength (low $q$) excitations, see Fig.~\ref{fig:stability}. In this regime the excitations spectrum has finite real and imaginary components and reads $\omega_{pq}\approx q\alpha(p) +i\beta(q)$ (the origin of this form will be clear from the discussion of the $s=0$ case here below).
These modes are not sufficient to produce a site-to-site dephasing, but are rather expected to induce collective oscillation of the system, as will be shown later.
iii)
Negative effective mass and DI appear in separate regions of the $p$-$q$ plane, thus opening the possibility for being investigated in a more clear way. For example,
one can tune the dispersion relation to negative values by loading the condensate in a quasimomentum state above $q_B/2$, avoiding the disturbance of DI that prevent a close inspection near the band edge in the repulsive case \cite{fallani03,fallani04}.

Following \cite{wuniu} we can compare these numerical results with
the analytics in two special cases: vanishing lattice ($s=0$) and noninteracting 
limit ($g=0$). For $s=0$ the 
diagonalization of the matrix $M$ yields the following analytic expression
for the excitation spectrum, $\omega_\pm=pq\pm\sqrt{q^4/4+cq^2}$, with $c=(3\sigma^2/2-1/\sigma^2
-1/2\sigma^2)/2$. This expression for $\omega_\pm$ is formally the same of the strictly 1D case \cite{wuniu}, 
but with a functional dependence of the interaction term $c$ on the condensate density $|\phi_p|^2$
(through $\sigma$). In case of attractive interactions, $g<0$, it is straightforward to verify that $c$ is always negative. In this case complex frequencies appear for $q^2<2|c|$ (for any $p$) and for 
$p=(\sqrt{(q-2q_B)^4/4+c(q-2q_B)^2}-\sqrt{q^4/4+cq^2})/2q_B$, as indicated in Fig. \ref{fig:stability}a-c by dashed and solid lines respectively.
In the former case, that is peculiar of attractive interactions, both real and imaginary component of $\omega_{pq}$ are finite, whereas in the latter the frequencies are purely imaginary (as for repulsive condensates \cite{wuniu,modugno}).
It is easy to verify that the former condition ($q^2<2|c|$) corresponds also to the 
the presence of negative eigenvalue of $\sigma_3 M$ even at $p=0$, the signature of a Landau instability. 

In the noninteracting limit $c=0$ the excitation energies can be directly obtained
from the noninteracting spectrum of the condensate $E(p)$, $\omega_\pm(q)=\pm(E(p\pm q)-E(p))$.
In this case the resonance condition reads $E(p+q)-E(p)=E(p)-E(p-q)$ \cite{wuniu}.
The points that satisfy the above constraint are shown by the line in Fig. \ref{fig:stability}g-i, and nicely reproduce the boundary between the unstable regions
for the attractive and repulsive cases.
For low $q$'s this resonance condition can be related to the condition for the vanishing of the
inverse effective mass $1/m^*$: in this case it can be easily demonstrated that besides the trivial solution $E_p=E_p$, one also has $q^2=12 E^{(2)}(p)/E^{(4)}(p)$ (where $E^{(n)}(p)=d^n E/dp^n$) leading to $1/m^{*}=0$ in the limit $q=0$.
This result provides an explanation of the numerical coincidence found in \cite{menotti}
concerning this point.

\begin{figure}
\centerline{\includegraphics[width=0.85\columnwidth,clip=]{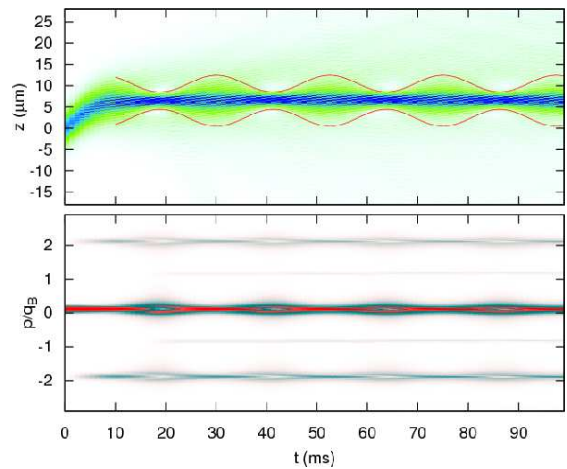}}
\caption{(Color online) Density plot of the condensate wavefunction along the axial direction
(top) and of its momentum distribution (bottom) as a function of time during the expansion in the waveguide, in the presence of a lattice moving at $v=0.2 v_B$ (see text).}
\label{fig:dynamics}
\end{figure}
To illustrate how the features of the spectrum discussed so far affect the system dynamics we now discuss the expansion of the condensate in a waveguide in the presence of a moving lattice. 
As an example we consider an elongated condensate composed by $N=1000$ atoms of $^{39}K$ with attractive interactions characterized by a scattering length $a=-a_0$, tuned by means of Feshbach resonances \cite{39K}. The BEC is initially prepared in a waveguide of radial frequency $\omega_r=2\pi\times1.5$ kHz, in the presence of an additional harmonic confinement of frequency $\omega_z=2\pi\times20$ Hz along the waveguide. 

Then, at $t=0$, the axial confinement is released and the condensate is let expand in the waveguide while a optical lattice of wavelength $\lambda=1040$ nm, moving at velocity $v$, is ramped up to a final intensity $s=5$ that is kept constant for the subsequent dynamics. As ramp time we choose $t_{ramp}=10$ ms, that is sufficient to project adiabatically the condensate in a state of quasimomentum $p/p_B\sim v/v_B$ \cite{fallani04,fallani03}.
This scheme is similar to experimental setup in \cite{fallani03}, with the difference that there also the radial confinement was removed in order to reach rapidly the non-interacting regime. Conversely, since here we want to probe the effects of nonlinearity, the presence of the waveguide helps to keep interactions for longer times.

The dynamics is investigated by solving the time dependent NPSE in Eq. (\ref{eq:npse}) for different velocities of the lattice \footnote{We recall that since GP-like equations are conservative they can account for DI but not for Landau instability \cite{modugno}. 
The latter is however characterized by longer timescales and can be safely neglected for the present discussion \cite{fallani04}.}.
Let us first consider the case $v=0.2 v_B$ shown in  Fig. \ref{fig:dynamics}, 
where we plot the evolution of the axial density and the momentum distribution in the  reference frame co-moving with the condensate at the end of the ramp.
According to the above linear stability analysis we expect the system to be 
dynamically unstable via long wavelength phonons. 
In particular, since the condensate occupies about $30$
sites with an average number of atoms per site $\bar{N}\approx 30$, the stability behaviour should be close to that shown in Fig. \ref{fig:stability}g.
Indeed the simulations show a rapid population of modes at small quasimomenta around $\bar{q}=0.058q_B$, corresponding to density modulations over several sites of the lattice, and matching the most unstable modes of the uniform system in Fig. \ref{fig:stability}g. 
Remarkably, the growth of these modes does not induce decoherence as observed so far with repulsive BECs \cite{fallani04}, but produces instead an oscillatory behaviour that is maintained for several periods. 
These oscillation can be accounted for by the finite real component of the excitations spectrum, and by taking into account the momentum spread $\Delta p$ due to finite size of the system.
Indeed, the fitted frequency $\nu=44.5$ Hz in Fig. \ref{fig:dynamics} is consistent with the
real part of frequency of the most unstable modes at $\bar{q}$.
Such a behaviour is conserved until the lattice velocity is tuned close to $v\simeq 0.5 v_B$, where the presence of unstable modes with $q\approx q_B$ favor a change of the condensate phase between neighboring sites of the lattice, leading to a complex dynamics that rapidly spoils the phase coherence of the BEC. 

\begin{figure}
\centerline{\includegraphics[width=0.85\columnwidth,clip=]{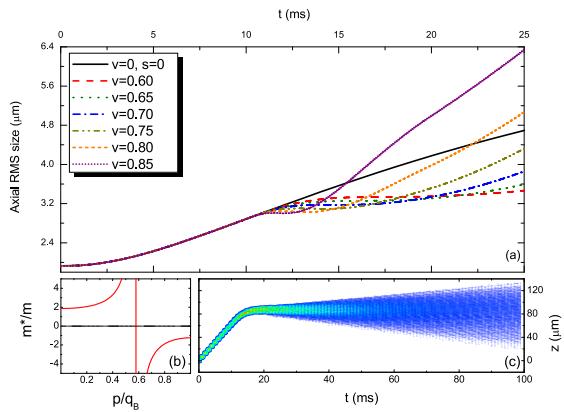}}
\caption{(Color online) (a) Evolution of the axial rms size of the condensate in the regime of negative effective mass, for different velocities, compared to the free expansion in the waveguide (solid line). The condensate is let expand freely for $10$ ms in the waveguide before ramping on the lattice. (b) Effective mass \textit{vs} quasimomentum. (c) Axial density plot of the expanding BEC for $v=0.8 v_B$.}
\label{fig:fig4}
\end{figure}
Another interesting scenario shows up in the dynamically stable region for $q>0.5 q_B$ where the the effective mass $m^*$ changes sign.
As discussed and experimentally demonstrated in \cite{fallani03,eiermann} a change of sign of $m^*$ corresponds to a time-reversed evolution and is therefore expected to produce a contraction of the axial size of a condensate initially expanding outwards \cite{massignan}. 
This is indeed shown in Fig. \ref{fig:fig4}(a) where we plot the evolution of the axial rms size for different velocities. This picture shows that in the presence of the lattice the expansion is slowed down until the wave packet eventually reaches its minimum allowed size $Z_{min}$, and then the dynamics inverts again. This effect is more evident for lower velocities where a larger negative $m^*$ (see Fig. \ref{fig:fig4}(b)) produces a hampering of the expansion for a longer time. Conversely, the expansion is enhanced near the band edge due to the lower value of $|m^*|$.
Notably this enhancement is favored also by the attractive interactions that are turned into an effective repulsion when $m^*$ changes sign, causing a boost of the expansion with respect
to the free case (even when $m<|m^*|$). This is the same mechanism that limits the value of $Z_{min}$ compared the noninteracting case \cite{massignan}.
Finally, Fig. \ref{fig:fig4}(c) demonstrates that in this regime of negative effective mass the system is not affected by DI, as discussed above.


In conclusion, we have discussed the occurrence of dynamical instability for attractive BECs moving in a 1D optical lattice in the presence of transverse confinement finding a remarkable behaviour with respect the usual repulsive case. Contrarily to the latter, attractive condensates are dynamically unstable at low quasimomenta and can be stabilized above a certain critical velocity. By suitably adjusting the lattice height and velocity it is also possible to tune
the disrupting character of dynamical instability \cite{fallani04} into a breathing-like collective oscillation of the system induced by long wavelength phonons.
Moreover, since negative effective mass and dynamical instability appear in separate regions of the $p$-$q$ plane of the condensate and excitations quasimomenta, this allows for a more clear investigation of both effects. 
We have also explicitly verified that the results presented here by means of the NPSE model
hold even in a full 3D GPE approach, provided that the radial confinement is tight enough to prevent the mixing of the radial branches of the spectrum (as the case considered here) \cite{modugno}. 
These results may be particularly relevant for current experiments with Bose-Einstein condensate with interactions tunable around zero \cite{lithium,cesium,39K}.

\begin{acknowledgments}
We thank L. Fallani and F. Dalfovo for the reading of the manuscript, and all the people of the QDG group at LENS for fruitful discussions. 
The work has been supported by MIUR, EU under contract MEIF-CT-2004-009939, and Ente CRF.
\end{acknowledgments}



\begin{thebibliography}{}
\bibitem{morsch}
O. Morsch and M. K. Oberthaler, Rev. Mod. Phys. \textbf{78}, 179 (2006). 

\bibitem{wuniu}
B. Wu and Q. Niu, Phys.  Rev.  A \textbf{64}, 061603(R) (2001).

\bibitem{smerzi}
A. Smerzi, A. Trombettoni, P. G. Kevrekidis and A. R. Bishop, 
Phys. Rev.  Lett \textbf{89}, 170402 (2002).

\bibitem{cataliottinjp}
F.~S.~Cataliotti \textit{et al.},
New Journ. Phys. \textbf{5}, 71 (2003).

\bibitem{menotti}
C. Menotti \textit{et al.},
New J. Phys. \textbf{5}, 112 (2003).

\bibitem{scott}
R. G. Scott \textit{et al.},
Phys. Rev. Lett. \textbf{90}, 110404 (2003);
R. G. Scott \textit{et al.},
Phys. Rev. A \textbf{69} 033605 (2004).

\bibitem{pethick}
M. Machholm, C.~J.~Pethick, and H.~Smith, 
Phys.  Rev.  A \textbf{67}, 053613 (2003); 
M. Machholm, A.~Nicolin, C.~J.~Pethick, and H.~Smith,
Phys. Rev. A \textbf{69}, 043604 (2004).

\bibitem{fallani04}
L. Fallani  \textit{et al.},
Phys. Rev. Lett. \textbf{93}, 140406 (2004).

\bibitem{modugno}
 M. Modugno, C. Tozzo, and F. Dalfovo,
Phys. Rev. A \textbf{70}, 043625 (2004);  Phys. Rev. A \textbf{71}, 
019904(E) (2005).

\bibitem{campbell}
G. K. Campbell \textit{et al.},
Phys. Rev. Lett. \textbf{96} 020406 (2006).

\bibitem{massignan}
P. Massignan and M. Modugno, Phys. Rev. A \textbf{67}, 023614 (2003).

\bibitem{fallani03}
L. Fallani \textit{et al.},
Phys. Rev. Lett. \textbf{91}, 240405 (2003).

\bibitem{eiermann}
B. Eiermann \textit{et al.},
Phys. Rev. Lett. 91, 060402 (2003).

\bibitem{lithium}
L. Khaykovich \textit{et al.},
Science 296, 1290 (2002); 
K. E. Strecker \textit{et al.},
Nature 417, 150 (2002).

\bibitem{cesium}
T. Weber \textit{et al.},
Science 299, 232 (2003).

\bibitem{39K}
G. Roati \textit{et al.}
cond-mat/0703714 (2007).

\bibitem{stringari}
S. Stringari and L. Pitaevskii, 
\textit{Bose-Einstein Condensation} (Oxford University Press, Oxford, 2003).

\bibitem{salasnich}
L. Salasnich, A. Parola and L. Reatto, 
Phys.  Rev.  A \textbf{65}, 043614 (2002).


\end{thebibliography}
\end{document}